\documentclass{amsart}

\usepackage{natbib}
\bibpunct{(}{)}{;}{a}{,}{,}

\usepackage{epsfig}

\newcommand{\Ei}{\mathrm{\,Ei\,}}

\begin{document}

\title[Non-equilibrium frequency spectrum]{Non-equilibrium theory of the allele frequency spectrum}

\author{Steven N.\ Evans $^1$}
\address{Steven N.\ Evans \\
  Department of Statistics \#3860 \\
  University of California at Berkeley \\
  367 Evans Hall \\
  Berkeley, CA 94720-3860 \\
  U.S.A.}
\email{evans@stat.Berkeley.EDU}
\urladdr{http://www.stat.berkeley.edu/users/evans/}
\thanks{(1) supported in part by NSF grant DMS-0405778.}

\author{Yelena Shvets $^2$}
\address{Yelena Shvets\\
Department of Statistics \#3860 \\
  University of California at Berkeley \\
  367 Evans Hall \\
  Berkeley, CA 94720-3860 \\
  U.S.A.}
\email{yelenashvets@yahoo.com}
\thanks{(2) Supported in part by NSF grant DMS-0405778.}

\author{Montgomery Slatkin $^3$}
\address{Montgomery Slatkin \\
Department of Integrative Biology \\
University of California at Berkeley \\
Berkeley, CA 94720-3140}
\email{slatkin@Berkeley.EDU}
\urladdr{http://ib.berkeley.edu/labs/slatkin/}
\thanks{(3) Supported in part by NIH grant R01-GM40282. Corresponding author.}





\begin{abstract}
A forward diffusion equation describing the 
evolution of the allele frequency spectrum is presented. 
The influx of mutations is accounted 
for by imposing a suitable boundary condition.
For a Wright-Fisher diffusion with or without
selection and varying population size, the boundary
condition is 
$\lim_{x \downarrow 0} 
x f(x,t)=\theta \rho(t)$, where $f(\cdot,t)$ is the 
frequency spectrum of derived alleles at 
independent loci at time $t$ and $\rho(t)$
is the relative population size at time $t$. When population 
size and selection intensity are independent of 
time, the forward equation is equivalent to the 
backwards diffusion usually used to derive
the frequency spectrum, but this approach 
allows computation of the time dependence of the 
spectrum both before an equilibrium is attained 
and when population size and selection intensity 
vary with time. From the diffusion equation,
a set of ordinary differential equations 
for the moments of $f(\cdot,t)$ is derived and the 
expected spectrum of a finite sample 
is expressed in terms of those 
moments.  The use of the forward 
equation is illustrated by considering neutral and selected 
alleles in a highly simplified model of human 
history.  For example, it is shown that approximately 30\% of the 
expected total heterozygosity of neutral loci is attributable to 
mutations that arose since the onset of 
population growth in roughly the last $150,000$ 
years.  
\end{abstract}

\maketitle

\section{Introduction}

The allele-frequency spectrum is the distribution 
of allele frequencies at a large number of
equivalent loci. 
The term ``site-frequency 
spectrum'' \citep{1}, is equivalent 
but emphasizes the application to individual 
nucleotides rather than alleles at different 
genetic loci. Here, we will use the frequency 
spectrum for both terms.

Although models assuming reversible mutation 
predict an equilibrium distribution of allele 
frequencies \citep{2}, all recent studies of 
frequency spectra assume irreversible mutation. 
Under that assumption, an equilibrium is not 
reached at any locus, but the distribution across 
polymorphic loci reaches an equilibrium if both 
population size and selection intensities are 
constant.  The theory predicting the frequency 
spectrum under irreversible mutation was 
developed by \citet{3}, \cite{4}, and \citet{5}. 
\citet{6} noted that this 
theory was applicable to nucleotide positions and 
introduced the ``infinitely many sites model.'' \citet{7} 
incorporated the theory of Fisher, 
Wright and Kimura into a  Poisson 
random field model for the purpose of estimating the 
selection intensity from the observed frequency 
spectrum in a finite sample of chromosomes. Their 
method has been tested and refined by 
\citet{8} and \citet{9}.

Past population growth affects the frequency 
spectrum.  \citet{12} showed that rapid 
growth resulted in more low frequency alleles 
than expected under neutrality.  \citet{11} 
confirmed that conclusion and examined the effect 
of past population growth on other aspects of the 
frequency spectrum. \citet{13} 
developed the coalescent theory for the frequency 
spectrum of neutral alleles in a population that 
has experienced arbitrary changes in population 
size. \citet{14} implemented the \citet{13} simulation method and applied 
it to human SNP data for the purpose of 
estimating the growth rate of human populations. 
Although \citet{14} was not able to reject 
the hypothesis of no growth, he noted that his 
analysis was of a small data set. \citet{15} considered the same problem and developed 
a method that allowed for both ascertainment bias 
and population subdivision. \citet{15} 
analyzed a larger SNP data set and found evidence 
of recent growth. \citet{16} and 
\citet{17} also modeled the 
coalescent process underlying the spectrum of neutral 
alleles and developed analytic theory that allows 
for exact calculation of the spectrum for large 
sample sizes.

\citet{18} reviewed and extended the theory 
of the frequency spectrum, making
clear the role of time-reversal.
He generalized that theory in two ways. 
He showed that the spectrum in a finite sample 
could be obtained from the solution to a 
backwards equation by assuming sampling with 
replacement, and he showed that the frequency 
spectrum in a population of variable size could 
be derived from the spectrum for a population of 
constant size when a transformation of the time 
scale reduces the backwards equation to one for a 
population of constant size.  The transformation 
of time scales is always possible for neutral 
alleles, in which case the frequency spectrum in 
a finite population is the same as that derived by 
\citet{13}.  For selected alleles, the 
frequency spectrum cannot be obtained by the method of
\citet{18} except in the special 
case in which the selection intensity is 
inversely proportional to the population size at 
all times in the past.

The frequency spectrum of alleles closely linked 
to selected loci is also of interest.  \citet{1}, \citet{19}, \citet{20},
 and others have simulated the effects of 
selected sites on the frequency spectrum of 
closely linked neutral sites, with the goal of 
finding evidence of background selection against 
deleterious mutations and genetic hitchhiking 
caused by positive selection of advantageous 
mutations.

\citet{10} recently considered the 
combined effects of population growth and 
selection on the frequency spectrum.  Their model 
was of a population that was of a constant size 
until $\tau$ generations in the past, at which 
time it grew instantaneously by a factor $\nu$ 
and remained at the new size until the present. 
\citet{10} developed a likelihood 
method for estimating both $\tau$ and $\nu$ from 
the spectrum of sites assumed to be neutral and 
for estimating $\tau$, $\nu$ and a scaled 
selection intensity $\gamma$ for non-neutral 
sites.  Their method was based on numerical 
solutions for the frequency spectrum using both 
\citet{21} series solution for neutral 
alleles and numerical solutions to the backwards 
equation for selected alleles.  
\citet{10} applied their method to a previously 
published data set of $301$ genes in the human 
genome and found evidence both of population 
growth and of purifying selection at 
non-synonymous sites.  In a related study, 
\citet{22} found evidence of 
differences in selection intensity among 
different classes of genes in a data set for more 
than $6000$ loci in humans.

In this paper, we will explore in more detail the 
allele frequency spectrum in a population of 
variable size.  Our goal is similar to that of 
\citet{10} in modeling the combined 
effects of selection and population growth.  We 
derive the frequency spectrum from the forward 
equation, first for a Markov chain and then for a 
diffusion approximation to that Markov chain. 
The forward equation provides a natural way to
compute the spectrum as it approaches an 
equilibrium from an arbitrary initial condition 
and to model the time-dependence of the spectrum 
resulting from the time-dependence of population 
size and selection intensity.  Furthermore, the 
forward equation provides a way to incorporate 
the effects of immigration. We show that
our approach recovers known results for an
equilibrium population
and present some numerical results for an 
idealized model of recent human populations.

\section{Markov chain}

The model is of a monoecious randomly-mating 
diploid population containing $N(t)$ individuals 
at time $t$, which in this section takes integer 
values representing discrete non-overlapping 
generations. We assume a large number of identical 
and independent loci.  At each locus there are 
only two alleles {\bf A}, the derived allele, and {\bf a}, 
the ancestral allele.  In generation $t$, the 
set of loci  is described by the row 
vector with $j^{\mathrm{th}}$ element $f_j(t)$ that is the expected  
numbers of loci at which {\bf A} is found on $j$ 
chromosomes, $1 \le j \le 2N(t)$.  Thus $f_{2N(t)}(t)$ is 
the expected 
number of loci fixed for {\bf A} in generation $t$. 
The model assumes that the pool of loci fixed for 
{\bf a} is so large that it can be assumed to be not 
reduced by the creation of polymorphic loci by 
mutation -- the infinitely many sites model of \citet{6}.

The change in $f_j(t)$ because of genetic drift 
and mutation is described by the set of 
difference equations
\begin{equation}
\label{eq:1}
f_j(t+1) = \sum_{i=1}^{2N(t)} f_i(t) p_{ij}(t)  + M_j(t),
\quad
1 \le j \le 2N(t+1).  
\end{equation}
The first term on the right 
hand side represents the combined effect of 
genetic drift and natural selection on loci that 
are already polymorphic: in the notation of \citet{23}, 
$p_{ij}(t)$ is the probability that a 
locus with $i$ copies of {\bf A} in generation $t$ will 
have $j$ copies in generation $t+1$.  The 
$p_{ij}(t)$ are easily derived for the 
Wright-Fisher and other models \citep{23}.

The second term on the right hand side represents 
the creation of new polymorphic loci by mutation 
and immigration.  The influx of mutations is 
modeled by assuming that each of the $2N(t)$ 
copies of {\bf a} at a monomorphic locus mutates to an 
{\bf A} with probability $\mu$ per generation.  Therefore
\begin{equation}
\label{eq:2}
M_j(t) = 2 N(t) \mu \delta_{1,j}
\end{equation}
for mutation alone, where $\delta_{1,j}=1$ if 
$j=1$ and $0$ otherwise.  Under the infinitely many 
sites model, the mutation rate is assumed to be 
so low that the effect of mutation on loci 
already polymorphic can be ignored.  However, 
mutation can be incorporated into $p_{ij}(t)$ if 
necessary.

Immigration from another population can be 
accounted for both by modifying $p_{ij}$ to allow 
for the effect of immigration on loci that are already 
polymorphic and $M_j(t)$ to allow for the creation of 
new polymorphic loci.  Immigration, unlike 
mutation, can create polymorphic loci for which 
$j>1$ immediately.  In this paper, we will 
restrict our analysis to the case with mutation 
only.

Given an initial frequency spectrum, $f_j(0)$, 
Equation (\ref{eq:1}) can be iterated to obtain the 
frequency spectrum at any time in the future. 
Even if $N$ is constant, there is no equilibrium 
solution to Equation (\ref{eq:1}) because the number of 
loci fixed for {\bf A} will increase each generation. 
However, an equilibrium solution for $f_j$, 
$1\le i \le 2N-1$, the spectrum for polymorphic loci, 
does exist and is found by solving the matrix 
equation
\begin{equation}
\label{eq:3}
\hat f = \hat f P + \frac{\theta}{2} e
\end{equation}
where $P$ is the $(2N-1)\times (2N-1)$ matrix with 
elements $p_{ij}$ for $1 \le i, j <2N$, $e$ is a row 
vector with the first element $1$ and the 
remaining elements $0$, and $\theta=4N\mu$.  The 
solution is
\begin{equation}
\label{eq:4}
\hat f = \frac{\theta}{2} e(I-P)^{-1},
\end{equation}
where $I$ is the $(2N-1) \times (2N-1)$ identity matrix.

This result is equivalent to that obtained using 
the backwards equation for the Markov chain.  The 
frequency spectrum of polymorphic loci is 
proportional to the sojourn times 
($\bar{t}_{1,j}$ in the notation of \citet{23}).  
Equation (\ref{eq:4}) is obtained from the 
solution to Equation (2.143) of \citet{23} by 
multiplying by  $\theta/2$, which is the rate 
of influx of mutations each generation.

The advantage of the formulation presented here 
is that it also describes the approach to the 
equilibrium.  The rate of approach depends on the 
second largest eigenvalue of $P$, which for a 
neutral allele is $1-\frac{1}{2N}$ \citep{23}.
The mean number of loci fixed for {\bf A} increases by
\begin{equation}
\label{eq:5}
\sum_{j=1}^{2N-1} f_j(t) p_{j,2N} 
\end{equation}
per generation, which reduces to $2N \mu P_1$ at 
equilibrium, where $P_1$ is the probability of 
fixation of each mutant.

\section{Diffusion approximation and new boundary condition}

Let us start by considering the time-homogeneous case with no mutation from the ancestral type,
but where we can start at time $0$ with some derived alleles 
already present.
Because we want to eventually allow varying population sizes, 
assume that that the population is described by
a time-homogeneous 
Markov chain with state-space $\{0,1,\ldots,2N \rho\}$.  Suppose
that if we shrink 
space by a factor of $2N \rho$ and speed time up by a factor of $2N$,
then this chain converges to a diffusion process on 
$[0,1]$
with generator ${\mathcal G} = a(x) \frac{d}{dx} + \frac{1}{2} b(x) \frac{d^2}{dx^2}$ 
for appropriate coefficients (this is the scaling regime that is
appropriate for models such as the Wright-Fisher chain with or
without selection).  

Suppose at time $0$ that there are
countably many loci at which derived alleles are present, 
with respective (non-random) frequencies 
$x_1, x_2, \ldots$.  
Once we have passed to the diffusion limit, the frequency spectrum
at time $t$ is just
the intensity measure (that is, the expectation measure) 
of the point process that
comes from starting independent copies 
of the diffusion process at each of the $x_i$
and letting them run to time $t$.  In other words,
the intensity measure is obtained by taking the sum of  point masses and moving it forwards an amount of time $t$ using the 
semigroup associated with the generator ${\mathcal G}$.

More generally, if the initial configuration of frequencies
is random (so that it can be thought of as a point process
on $(0,1)$), then the frequency spectrum at time $t$
is obtained by taking the measure that is the intensity
of that point process and again moving it forwards an amount
of time $t$ using this semigroup. 

For $t>0$ the resulting measure will be absolutely continuous
with respect to Lebesgue measure and have
a density $f^o(y,t)$ at frequency $y \in (0,1)$.
We will also refer to this density as the
frequency spectrum. It is immediate that
$f^o$ satisfies the Kolmogorov forward equation
equations that go with the generator $\mathcal G$ (with initial conditions
corresponding to the intensity measure of the point process
of initial frequencies).
That is,
\begin{equation}
\frac{\partial}{\partial t} f^o(y,t) = - \frac{\partial}{\partial y} [a(y) f^o(y,t)] + \frac{1}{2} \frac{\partial^2}{\partial y^2} [b(y) f^o(y,t)],
\end{equation}
with
$\lim_{y \downarrow 0} f^o(y,t)$ and $\lim_{y \uparrow 1} f^o(y,t)$
both finite and
appropriate boundary conditions at $t=0$ (in particular,
if the point process of initial frequencies has intensity
$h(y) \, dy$, then $\lim_{t \downarrow 0} f^o(y,t) = h(y)$).

Now we want to introduce mutation from the ancestral type 
as time progresses.  In the Markov chain model,
this corresponds to new mutants arising in the population
at rate $\frac{\theta}{2} \rho$ per unit of Markov chain time,
where $\theta$ is independent of $N$.  
The initial number of mutants at a locus is $1$.
This is equivalent to mutants appearing at rate
$2N \frac{\theta}{2} \rho$ per unit of rescaled time, with the
initial proportion of mutants at a locus being $\frac{1}{2N\rho}$.

Imagine now that we pass to the diffusion limit
for the allele frequencies, but  for the moment still work with a finite
$N$ for the description of the appearance of new mutants.  
That is, we think of our evolving point process as having
new points added at location $\frac{1}{2N\rho}$ at rate 
$\frac{\theta}{2} 2N \rho$, and that the locations
of these points then evolve as independent copies of the diffusion with generator ${\mathcal G}$.

We will make substantial use of the theory of entrance laws for
one-dimensional diffusions laid out in Section 3 of \citet{MR656509}.
Write $P_t(x,dy)$ for the semigroup associated with ${\mathcal G}$.
This is the semigroup of the {\em $0$-diffusion} in the terminology of \citet{MR656509}.
The contribution to the frequency spectrum from mutations that
appear after time $0$ is
\begin{equation}
2 N \frac{\theta}{2}  \rho \int_0^t P_{t-s}\left(\frac{1}{2 N \rho},dy\right) \, ds.
\end{equation}

Choose a scale function $s$ for the $0$-diffusion such that $s(0) = 0$
(so that $s$ is then unique up to a positive multiple).  
As  \citet{MR656509} observe, 
\begin{equation}
P_u^{\uparrow}(x,dy) := \frac{1}{s(x)} P_u(x,dy) s(y), 
\quad 0 < x,y \le 1,
\end{equation}
is the semigroup of a diffusion that never hits $0$ (this
{\em $\uparrow$-diffusion} is the Doob $h$-transform
that corresponds to the naive idea of conditioning the $0$-diffusion never to hit $0$).  
Moreover, this semigroup
can be extended to allow starting at $0$
by setting
\begin{equation}
P_u^{\uparrow}(0,dy) = \lim_{x \downarrow 0} P_u^{\uparrow}(x,dy).
\end{equation}
The resulting extended process can start at $0$ but it
will never return to $0$.

Assume now that $s'(0)>0$, which will be the case in the diffusions
that are of interest to us.
We can choose the free multiplicative
constant in the definition of the scale function $s$
so that $\lim_{y \downarrow 0} s(y)/y = s'(0) = 1$.  Then
\begin{equation}
\lim_{N \rightarrow \infty} 2 N \rho P_u\left(\frac{1}{2N \rho},dy\right) = \frac{P_u^{\uparrow}(0,dy)}{s(y)} =: \lambda_u(dy)
\end{equation}
in the notation of Pitman and Yor.  Thus
\begin{equation}
\lim_{N \rightarrow \infty} 2 N \frac{\theta}{2} \rho \int_0^t P_{t-s}\left(\frac{1}{2 N \rho},dy\right) \, ds 
= \frac{\theta}{2}  \int_0^t \lambda_{t-s}(dy) \, ds
=: \Phi_t(dy),
\end{equation}
say.

As observed in \citet{MR656509}, 
the family $(\lambda_u)_{u > 0}$ is an entrance law for the semigroup
of the $0$-diffusion, and so it has  densities  that satisfy the Kolmogorov forward equation associated with
the generator ${\mathcal G}$ --- intuitively, $(\lambda_u)_{u > 0}$ describes that injection of an infinite amount
of mass at location $0$ at time $0$, with this mass subsequently evolving in $(0,1)$ according to the dynamics of the $0$-diffusion.  Consequently,
the family $(\Phi_t)_{t>0}$ also  satisfies the Kolmogorov forward equation associated with 
the generator ${\mathcal G}$ --- again intuitively, $(\Phi_t)_{t>0}$ describes a continuous--in--time injection of mass at location $0$,
with this mass again subsequently evolving in $(0,1)$ according to the dynamics of the $0$-diffusion.  That is, if we write $\phi_t$
for the density of $\Phi_t$, we have that 
\begin{equation}
\frac{\partial}{\partial t} \phi_t(y) = - \frac{\partial}{\partial y} [a(y) \phi_t(y)] + \frac{1}{2} \frac{\partial^2}{\partial y^2} [b(y) \phi_t(y)].
\end{equation}

It remains to work out what the boundary conditions for $\phi_t$ are.
Following \citet{MR656509}, introduce the {\em $\downarrow$-diffusion},
which is the $0$-diffusion conditioned to hit $0$ before $1$.
The $\downarrow$-diffusion has the Doob $h$-transform semigroup
\begin{equation}
P^\downarrow_t(x,dy) = 
\left(1 - \frac{s(x)}{s(1)}\right)^{-1}
P_t(x,dy)
\left(1 - \frac{s(y)}{s(1)}\right).
\end{equation}
From \citet{MR0350881}, the $\uparrow$-diffusion started
at $0$ and killed at the last time it visits $y>0$ is the
time-reversal of the $\downarrow$-diffusion started 
at $y$ and killed when it first hits $0$.  Write
$(Q_t^\downarrow)_{t \ge 0}$ for the semigroup of this killed process.
Since we have normalized the scale function $s$
so that $s(y) \approx y$ for $y$ close to $0$, 
\begin{equation}
\begin{split}
\lim_{y \downarrow 0} y \phi_t(y)
& =
\lim_{y \downarrow 0} s(y) \phi_t(y) \\
& =
\lim_{y \downarrow 0}
 s(y) \int_0^t \frac{\theta}{2} \frac{\lambda_{t-s}(dy)}{dy} \, ds  \\
& =
\frac{\theta}{2} \lim_{y \downarrow 0}
\int_0^t \frac{P_{t-s}^\uparrow(0,dy)}{dy} \, ds \\
& = \frac{\theta}{2} \lim_{y \downarrow 0}
\int_0^t \frac{P_s^\uparrow(0,dy)}{dy} \, ds \\
& = \frac{\theta}{2} \lim_{y \downarrow 0}
\int_0^\infty \frac{P_s^\uparrow(0,dy)}{dy} \, ds \\
& = \frac{\theta}{2} \lim_{y \downarrow 0}
\int_0^\infty \frac{Q_s^\downarrow(y,dy)}{dy} \, ds. \\
\end{split}
\end{equation}

Note that if $(B_t)_{t \ge 0}$ is a standard Brownian motion
and $T := \inf\{t \ge 0 : B_t = 0\}$, then, by Equation (3.2.1)
and Section 3.1 of \citet{MR613983},
and
\begin{equation}
\begin{split}
\int_0^\infty \frac{\mathbb{P}^y \{B_s \in dy, T > s\}}{dy} \, ds
& =
\int_0^\infty \frac{1}{\sqrt{2 \pi s}} - \frac{1}{\sqrt{2 \pi s}} e^{-(2y)^2/2s} \, ds \\
& = \lim_{\lambda \downarrow 0} \frac{1}{\sqrt{2 \lambda}}
- \frac{1}{\sqrt{2 \lambda}} \exp(- \sqrt{2 \lambda} 2 y) \\
& = 2 y. \\
\end{split}
\end{equation}
Observe also that a scale function for the $\downarrow$-diffusion
is 
\begin{equation}
\sigma(x) := \frac{s(x) s(1)}{s(1) - s(x)}.
\end{equation}
By standard one-dimensional diffusion theory, if we compose the killed
$\downarrow$-diffusion with $\sigma$, then the resulting process
is a time-change of standard Brownian motion killed when it
first hits $0$, with the time-change given by the corresponding
speed measure (see, for example, V.7 of \citet{MR1780932}).
Moreover, since $\sigma(x) \sim x$ for $x$ close to $0$, the
speed measure $m$ for the $\downarrow$-diffusion satisfies
$m(dx) \sim dx / b(x)$ for $x$ close to $0$ (beware that
the definitions of the speed measure can vary from author to author
by multiplicative constants, we are using the definition
of \citet{MR1780932}).  Therefore
\begin{equation}
\frac{\theta}{2} \lim_{y \downarrow 0}
\int_0^\infty \frac{Q_s^\downarrow(y,dy)}{dy} \, ds
= \frac{\theta}{2} \lim_{y \downarrow 0} \frac{2 y}{b(y)}
= \theta \lim_{y \downarrow 0} \frac{y}{b(y)}.
\end{equation}

Write $f(x,t)$ for the frequency spectrum of the model with mutation
from ancestral type.
We have $f(x,t) = f^o(x,t) + \phi_t(x)$, where $f^o$ is defined for the
appropriate initial conditions at $t=0$.  If we want to start with all
alleles ancestral type, then the initial conditions at $t=0$
are null and $f^o \equiv 0$.
Combining what we have
obtained above, we find that
\begin{equation}
\frac{\partial}{\partial t} f(x,t) = - \frac{\partial}{\partial x} [a(x) f(x,t)] + \frac{1}{2} \frac{\partial^2}{\partial x^2} [b(x) f(x,t)].
\end{equation}
with appropriate boundary conditions at $t=0$ (in particular,
$\lim_{t \downarrow 0} f(x,t) = 0$ if we start will all alleles being ancestral),
and further boundary conditions $\lim_{x \downarrow 0} x f(x,t) = \theta
\lim_{x \downarrow 0} \frac{x}{b(x)}$ and $\lim_{x \uparrow 1} f(x,t)$ 
finite.

Now consider a time-inhomogeneous diffusion with generator
$a(x,t) \frac{d}{dx} + \frac{1}{2} b(x,t) \frac{d^2}{dx^2}$ and suppose also that $\rho$ is now a function $\rho(t)$ of time.  By 
first considering the
case where $a$, $b$ and $\rho$ are piecewise constant, using the above analysis, and then taking limits,
we get that the frequency spectrum solves
\begin{equation}
\label{eq:PDE_for_f}
\frac{\partial}{\partial t} f(y,t) = - \frac{\partial}{\partial y} [a(y,t) f(y,t)] + \frac{1}{2} \frac{\partial^2}{\partial y^2} [b(y,t) f(y,t)]
\end{equation}
with appropriate boundary conditions at $t=0$ 
and further boundary conditions 
$\lim_{y \downarrow 0} y f(y,t) = \theta \lim_{x \downarrow 0} \frac{x}{b(x,t)}$ and 
$\lim_{y \uparrow 1} f(y,t)$ finite.

For the purposes of a numerical solution, it is more convenient
to consider the function $g(x,t) := x(1-x) f(x,t)$ which satisfies
\begin{equation}
\label{eq:PDE_for_g}
\frac{\partial}{\partial t} g(x,t)  =  
- x(1-x)\frac{\partial}{\partial x} \left[\frac{a(x,t)}{x(1-x)} g(x,t)\right] 
+ \frac{x(1-x)}{2} \frac{\partial^2}{\partial x^2} \left[\frac{b(x,t)}{x(1-x)} g(x,t)\right]
\end{equation}
with appropriate boundary conditions at $t=0$ 
and further boundary conditions 
$\lim_{x \downarrow 0}  g(x,t) = \theta \lim_{x \downarrow 0} \frac{x}{b(x,t)}$ and 
$\lim_{x \uparrow 1} g(x,t) = 0$.

As an example, consider the case where  $a(x) = S x(1-x)$, and
$b(x) = x(1-x)/\rho(t)$.  This is a Wright-Fisher diffusion
with selection and varying population size.  
The corresponding forward equation is
\begin{equation}
\label{eq:PDE_for_g_WF}
\frac{\partial}{\partial t} g(x,t)  =  
- S x(1-x) \frac{\partial}{\partial x} \left[ g(x,t)\right] 
+ \frac{x(1-x)}{2 \rho(t)} \frac{\partial^2}{\partial x^2} \left[g(x,t)\right]
\end{equation}
with boundary conditions
\begin{equation}
\lim_{x \downarrow 0}  g(x,t) 
= 
\theta \lim_{x \downarrow 0} \frac{x \rho(t)}{x(1-x)}
=
\theta \rho(t).
\end{equation}

\section{Equilibrium solution}

When $\rho(t) \equiv 1$ is a constant and the coefficients $a$ and $b$
do not depend on time, then we expect $f(\cdot, t)$ to converge
to an equilibrium $\hat f$ as $t \rightarrow \infty$.

From Equation (\ref{eq:PDE_for_f}), the function $\hat f$ should satisfy
\begin{equation}
\label{eq:PDE_for_hatf}
0 = - \frac{\partial}{\partial x} [a(x) \hat f(x)] + \frac{1}{2} \frac{\partial^2}{\partial x^2} [b(x) \hat f(x)]
\end{equation}
with boundary conditions 
$\lim_{x \downarrow 0} x \hat f(x) = \theta \lim_{x \downarrow 0} \frac{x}{b(x)}$ and 
$\lim_{x \uparrow 1} f(x)$ finite.

In order to solve this equation, suppose first that the diffusion process is on
natural scale in $[0,1]$, so that $a \equiv 0$.  In that case,
Equation (\ref{eq:PDE_for_hatf}) becomes
\begin{equation}
0 =   \frac{1}{2} \frac{\partial^2}{\partial x^2} [b(x) \hat f(x)],
\end{equation}
so that $b(x) \hat f(x) = c_0 + c_1 x$ for some constants $c_0$
and $c_1$.
Assume that $0<\lim_{x \uparrow 1} (1-x)/b(x)<\infty$.
Satisfying the boundary conditions requires that 
$\hat f(x) = \theta (1-x)/b(x)$.

Suppose now that the diffusion process is not on natural scale
and that $s$ is a scale function with $s(0)=0$ and $s(1) = 1$,
so that the image of the diffusion under $s$ is a diffusion
on $[0,1]$ in natural scale.  The generator of the image
diffusion is
\begin{equation}
\frac{1}{2} [(s'\circ s^{-1})(x)]^2
(b \circ s^{-1})(x)\frac{\partial^2}{\partial x^2},
\end{equation}
and hence the image diffusion has the associated frequency spectrum
\begin{equation}
\hat h(x) := \frac{\theta (1-x)}{[(s'\circ s^{-1})(x)]^2 
(b \circ s^{-1})(x)}
\end{equation}
from what we have just observed.
It follows from the usual change of variable
formula for densities that the original diffusion has 
the associated frequency spectrum
\begin{equation}
\hat f(x) = (\hat h \circ s)(x) s'(x)
=
\frac{\theta (1 - s(x))}{s'(x) b(x)}.
\end{equation}

In terms of coefficients,
\begin{equation}
s(x) = 
\frac{
\int_0^x \exp\left(-2 \int_0^y a(z)/b(z) \, dz\right) \, dy
}
{
\int_0^1 \exp\left(-2 \int_0^y a(z)/b(z) \, dz\right) \, dy.
}
\end{equation}
Also,
\begin{equation}
\frac{1}{s'(x) b(x)} = \frac{d m}{dx}(x),
\end{equation}
where $m$ is the speed measure corresponding to the scale
measure $s$ (recall that we are using the normalization of
the speed measure in \citet{MR1780932}).  Thus
\begin{equation}
\hat f(x) = \theta P_0(x) \frac{d m}{dx}(x),
\end{equation}
where $P_0(x)$ is the probability the diffusion will be
absorbed at $0$ given that it starts at $x$. This well-known
equation can be established directly via an ergodic argument using time-reversal -- see \citet{18} for a discussion of the history
of this technique.  Note that our derivation of
Equation (\ref{eq:PDE_for_f}) also used time-reversal.

For example, when $a = 0$ and $b(x)=x(1-x)$ (the neutral
Wright-Fisher diffusion),  $\hat f(x) = \theta/x$, agreeing with
Equation (9.18) of \citet{23}.
Similarly, when $a(x) = S x(1-x)$ and $b(x)=x(1-x)$ (the 
Wright-Fisher diffusion with selection and no dominance), 
\begin{equation}
\hat f(x) = \theta \frac{e^{2S}\left(1 - e^{-2S(1-x)}\right)}
{(e^{2S}-1)x(1-x)},
\end{equation}
agreeing with Equation (9.23) of \citet{23}.

\section{A system of ODEs for the moments in a Wright-Fisher diffusion
with varying population size}

Suppose in this section that 
$a(x) = S x(1-x)$ and $b(x) = x(1-x)/\rho(t)$.
This is a Wright-Fisher diffusion model with
time-varying population size and
constant selection and without dominance.
Equation (\ref{eq:PDE_for_g_WF}) applies.

Put $\mu_n(t) := \int_0^1 x^n g(x,t) \, dx$ for $n=0,1,2,\ldots$.  
Integrating by parts, we get
\begin{equation}
\begin{split}
&\int_0^1 x^n x(1-x)\frac{\partial }{\partial x} g(x,t) \, dx \\
& \quad =
\left[(x^{n+1}-x^{n+2}) g(x,t) \right]_0^1 \\
& \qquad
-
\int_0^1 ((n+1)x^{n}-(n+2)x^{n+1}) g(x,t) \, dx \\
& \quad =
(n+1) \mu_n - (n+2) \mu_{n+1}(t) \\
\end{split}
\end{equation}

Similarly, 
\begin{equation}
\begin{split}
&\int_0^1 x^n x(1-x)\frac{\partial^2 }{\partial x^2} g(x,t) \, dx \\
& \quad =
\left[(x^{n+1}-x^{n+2})\frac{\partial }{\partial x} g(x,t) \right]_0^1 \\
& \qquad
-
\int_0^1 ((n+1)x^{n}-(n+2)x^{n+1}) \frac{\partial }{\partial x} g(x,t) \, dx \\
& \quad =
- \int_0^1 ((n+1)x^{n}-(n+2)x^{n+1}) \frac{\partial }{\partial x} g(x,t) \, dx \\
& \quad =
- \left[((n+1)x^{n}-(n+2)x^{n+1}) g(x,t) \right]_0^1 \\
& \qquad +
\int_0^1 ((n+1)n x^{n-1}1\{n \ne 0\} -(n+2)(n+1)x^{n}) g(x,t) \, dx \\
& \quad =
 \left[1\{n=0\} \theta \rho(t) \right]
+
\left[(n+1)n \mu_{n-1}(t)1\{n \ne 0\} - (n+2)(n+1) \mu_n(t) \right] \\
\end{split}
\end{equation}

We thus get the coupled system of ODEs
\begin{equation}
\mu_0'(t) = \frac{\theta}{2} - \frac{1}{\rho(t)} \mu_0(t)
+ S\left(\mu_0(t) - 2 \mu_1(t)\right)
\end{equation}
and
\begin{equation}
\begin{split}
\mu_n'(t) & = \frac{1}{2 \rho(t)}\left[(n+1)n \mu_{n-1}(t) - (n+2)(n+1) \mu_n(t) \right] \\
& \quad + S \left((n+1) \mu_n(t) - (n+2) \mu_{n+1}(t)\right), \quad n \ge 1. \\
\end{split}
\end{equation}

When $S=0$ (that is, the neutral case) this system is lower triangular and we
can be solved explicitly.  The ODE for $\mu_0$ has solution
\begin{equation}
\label{explicit0}
\mu_0(t) = 
\mu_0(0) \exp\left(-\int_0^t \frac{1}{\rho(s)} \, ds\right)
+
\frac{\theta}{2} \frac{\int_0^t \exp\left(\int_0^s \frac{1}{\rho(u)} \, du \right) \, ds} {\exp\left(\int_0^t \frac{1}{\rho(s)} \, ds\right)}.
\end{equation}
Given $\mu_{n-1}$, the ODE for $\mu_n$ has solution
\begin{equation}
\label{explicitn}
\begin{split}
\mu_n(t) & = 
\mu_n(0)\exp\left(-\binom{n+2}{2} \int_0^t \frac{1}{\rho(s)} \, ds\right) \\
& \quad +
\binom{n+1}{2} 
\frac{\int_0^t \frac{1}{\rho(s)} \mu_{n-1}(s) \exp\left(\binom{n+2}{2} \int_0^s \frac{1}{\rho(u)} \, du \right) \, ds} 
{\exp\left( \binom{n+2}{2} \int_0^t \frac{1}{\rho(s)} \, ds \right)}. \\
\end{split}
\end{equation}

We can draw some conclusions from these equations about the effect of
$\rho$ on the asymptotic behavior of $\mu_n$.  For example, 
recall that the measure $f(x,t) \, dx$ is the intensity of
the point process on $(0,1)$ that records the frequencies of
derived alleles at all the loci at which derived alleles are
present with non-zero frequency at time $t$.  Hence 
$2 \mu_0(t) = 2 \int_0^1 x(1-x) f(x,t) \, dx$ is the expected
value of the total of the heterozygosities summed over all loci.
For any
initial conditions and any
$\rho$ such that 
$\int_0^\infty \frac{1}{\rho(t)} \, dt < \infty$,
the expected total heterozygosity $2 \mu_0(t)$
is asymptotically equivalent to $\theta t$ as $t \rightarrow \infty$.

\section{Explicit recurrences for the moments in a 
Wright-Fisher diffusion with exponentially increasing population size}

Suppose in this section that $\rho(t) = e^{Rt}$ with $R>0$,
$a(x) = 0$, and $b(x) = x(1-x)/e^{Rt}$.  This
is a neutral Wright-Fisher diffusion model with
constant selection and exponentially increasing population size.
We will obtain an explicit recursive recipe for the moments $\mu_n(t)$.
For simplicity, suppose that $f(x,0) \equiv 0$, so that
$\mu_n = 0$ for all $n$.  A similar development holds
for other initial conditions. 

Recall that the {\em exponential integral function} $\Ei$ is given by
$\Ei(x) = -\int_{-x}^\infty t^{-1} e^{-t} \, dt$, where the principal
value is taken if $x>0$ (although we are only interested in the case $x<0$).
For $x<0$, $\Ei(x) = -\Gamma(0,-x)$, where $\Gamma$ is the usual upper
incomplete gamma function.  For $x>0$,
\begin{equation}
\begin{split}
\Ei(-x) & = \gamma + \log(x) + \int_0^x \frac{e^{-t} - 1}{t} \, dt \\
& = \gamma + \log(x) + \sum_{j=1}^\infty \frac{(-x)^j}{j \cdot j!}, \\
\end{split}
\end{equation}
where $\gamma$ is Euler's constant \citep{MR1773820}.

For $n=0$,
\begin{equation}
\mu_0(t)
=
\frac{\theta}{2R} e^{\frac{e^{-R t}}{R}}
   \left(\Ei\left(-\frac{1}{R}\right)-\Ei
   \left(-\frac{e^{-R t}}{R}\right)\right).
\end{equation}

For $n=1,2,\ldots$ define a linear operator $\Phi_n$ by
\begin{equation}
\Phi_n h(t) := \binom{n+1}{2} \frac{\int_0^t e^{-R s} h(s) \exp\left(\binom{n+2}{2} \int_0^s e^{-Ru} \, du \right) \, ds} {\exp\left( \binom{n+2}{2} \int_0^t e^{-Rs} \, ds \right)},
\end{equation}
so that $\mu_n = \Phi_n \mu_{n-1} = \ldots = \Phi_n \Phi_{n-1} \cdots \Phi_1 \mu_0$.
Set 
\begin{equation}
\begin{split}
h_k(t)& = \exp\left(\binom{k+2}{2}\frac{e^{-Rt}}{R}\right) \\ 
& \quad \times \left[
\Ei\left(-\binom{k+2}{2}\frac{1}{R}\right)
-
\Ei\left(-\binom{k+2}{2}\frac{e^{-Rt}}{R}\right)
\right], \\
\end{split}
\end{equation}
so that $\mu_0 = \frac{1}{2R} h_0$.
It follows from a straightforward integration that
\begin{equation}
\Phi_n h_k(t) =
\frac{\binom{n+1}{2}}{\binom{n+2}{2}-\binom{k+2}{2}}
[h_k(t) - h_n(t)].
\end{equation}
Hence
\begin{equation}
\mu_n(t) := \sum_{k=0}^n c_{n,k} h_k(t),
\end{equation}
where $c_{0,0} = \frac{1}{2R}$ and the other $c_{n,k}$ are given recursively by
\begin{equation}
c_{n,k} 
= \frac{\binom{n+1}{2}}{\binom{n+2}{2}-\binom{k+2}{2}} c_{n-1,k}, \quad 1 \le i \le n-1,
\end{equation}
and
\begin{equation}
\begin{split}
c_{n,n}
& =
- \sum_{k=0}^{n-1} 
\frac{\binom{n+1}{2}}{\binom{n+2}{2}-\binom{k+2}{2}}
c_{n-1,k} \\
& =
- \sum_{k=0}^{n-1} 
c_{n,k}. \\
\end{split}
\end{equation}

For example,
\begin{equation}
\begin{split}
c_{1,0} & = \frac{1}{3-1} \frac{1}{2R} = \frac{1}{4R} \\
c_{1,1} & = \frac{-1}{4R} \\
\end{split}
\end{equation}
and
\begin{equation}
\begin{split}
c_{2,0} & = \frac{3}{6-1} \frac{1}{4R} = \frac{3}{20R} \\
c_{2,1} & = \frac{3}{6-3} \frac{(-1)}{4R} = - \frac{1}{4R} \\
c_{2,2} & = -\left(\frac{3}{20R} -  \frac{1}{4R} \right) = \frac{1}{10R}.\\
\end{split}
\end{equation}

Set $\psi(x) := \sum_{j=1}^\infty \frac{x^j}{j \cdot j!}$.
  Then   
\begin{equation}
\begin{split}
h_k(t)
& = \exp\left(\binom{k+2}{2}\frac{e^{-Rt}}{R}\right) \\ 
& \quad \times \left[
\Ei\left(-\binom{k+2}{2}\frac{1}{R}\right)
-
\Ei\left(-\binom{k+2}{2}\frac{e^{-Rt}}{R}\right)
\right] \\
& = \exp\left(
\binom{k+2}{2}\frac{e^{-Rt}}{R}\right) \\
& \quad \times \left[Rt + \psi\left(-\binom{k+2}{2}\frac{1}{R}\right) 
- \psi\left(-\binom{k+2}{2}\frac{e^{-Rt}}{R}\right)
\right]. \\
\end{split}
\end{equation}  

It follows from this that
\begin{equation}
\mu_0(t) \approx \frac{\theta t}{2}
\end{equation}
when $R$ is large (recall that $2 \mu_0(t)$ is the expected total
heterozygosity summed over all loci). 
For $n \ge 1$, the two observations that $\exp\left(-\binom{k+2}{2}\frac{e^{-Rt}}{R}\right)$ will be very close to
$1$ for even moderate values of $R$ and that $\sum_k c_{n,k} = 0$ 
show that the contribution to $\mu_n(t)$ from the $Rt$ term will almost cancel out
and the primary contribution will be from the $\psi$ terms.

\section{Frequency spectrum in a finite sample}
\label{S:finite_sample}

The function $f(x,t)$ approximates the frequency 
spectrum in a very large population.  In a sample 
of $n$ chromosomes, we can observe only the 
finite spectrum, $f_i(t)$, which is the 
distribution of the number of chromosomes at 
which there are $i$ derived alleles ($0<i \le n$). In 
this context, $f_i(t)$ is similar to $f_i(t)$ 
defined for the Markov chain formulation, but 
here $t$ is continuous.  The finite spectrum is 
obtained from $f(x,t)$ by assuming sampling with 
replacement at each locus independently:
\begin{equation}
\label{eq:XX+1}
f_i(t) = \binom{n}{i} \int_0^1 x^i(1-x)^{n-i} f(x,t) \, dx
\end{equation}
\citet{18}.  We can express this equation 
in terms of the moments of $g(x,t)=x(1-x)f(x,t)$ 
about $x=0$:
\begin{equation}
\label{eq:XX+2}
f_i(t) = \binom{n}{i} \sum_{j=0}^{n-i-1} (-1)^j \binom{n-i-1}{j} \mu_{j+i-1}(t)
\end{equation}
This expression involves an alternating sum and so, from a numerical
point of view, it might be preferable to have 
a system of ODEs for the $f_i$
themselves rather than for the $\mu_i$.  Unfortunately, a
system of ODEs for the $f_i$ appears to be rather complicated:
it is doubly indexed by $i$ and the suppressed index $n$ and,
moreover,  
the alternation present in Equation (\ref{eq:XX+2}) is just ``transferred''
to the coefficients of the ODEs. 

For use later in Section \ref{S:hum_pop_growth} we make the following small observation.
Suppose that $x \mapsto f(x,t)$ is decreasing.  Observe for $1 \le i \le n-1$ that,
by an integration by parts,
\begin{equation}
\begin{split}
&f_{i+1}(t) - f_i(t) \\
& \quad =
\frac{n!}{(i+1)! (n-i)!} \int_0^1 [x^{i+1} (n-i) (1-x)^{n-i-1} - (i+1)x^i (1-x)^{n-i}] \, f(x,t) \, dx \\
& \quad =
- \frac{n!}{(i+1)! (n-i)!} \int_0^1 \frac{\partial}{\partial x} [x^{i+1} (1-x)^{n-i}] \, f(x,t) \, dx \\
& \quad =
\frac{n!}{(i+1)! (n-i)!} \int_0^1  x^{i+1} (1-x)^{n-i} \, \frac{\partial}{\partial x} f(x,t) \, dx \\
& \quad < 0 \\
\end{split}
\end{equation}
(there are no boundary terms because of the boundary conditions on $f(\cdot,t)$).
Thus $f_i(t)$ is decreasing in $i$ when $x \mapsto f(x,t)$ is decreasing.

\section{Numerical analysis}

If the population size or selection intensity 
vary only somewhat with time, numerical solutions 
to Equation (\ref{eq:PDE_for_g_WF})
can be obtained with standard methods.  The 
function {\tt NDSolve} in {\em Mathematica} (version 5) and 
the function {\tt pdepe} of {\em MATLAB} (version 7) both 
provide solutions using default settings for 
those programs.  With extreme population growth, 
as in the model of human history we 
consider in Section~\ref{S:hum_pop_growth}, neither of these programs provides 
accurate solutions, so it was necessary to write 
a program tailored to the problem.

Large gradients at the boundary $x=0$ make it necessary
to introduce a non-uniform grid with $N$ internal points. With the view toward integrating
the numerical solution in order to compute moments and the finite
spectrum we select the grid with the smallest spatial increment $\Delta_0 = (x_1-0)$
of the order $10^{-8}$ and the largest $\Delta_N = (1-x_N) \sim 10^{-3}$. 
The spacing is kept constant for a few nodes $x_i,...,x_{i+k}$ and then doubled so that 
$$x_{i + k+1} - x_{i+k}= \Delta_{i+k} =2\Delta_{i+k-1}= 2(x_{i + k} - x_{i+k-1}).$$
This process is repeated until the spacing reaches the maximum size $\Delta_N \sim 10^{-3}$.
The numerical domain is thus separated into sub-domains within each of which the spacing is uniform.
This guarantees that centered difference schemes used in the sub-domains 
give a  second order accurate approximations to the corresponding differential operators.
For the right end-point of the uniform  sub-domains, $x_{i+k}$, (referred to as the edge-node)  
the values at $x_{i+k-2}$ and $x_{i+k+1}$ can be used, since
$x_{i+k} - x_{i+k-2} = x_{i+k+1} - x_{i+k}$.  

If $t= \Delta_tk$ and $x=x_i$ we denote the numerical approximation to $g(x,t)$ by $G_i^k$.
A standard centered second order differencing is used to approximate
the diffusive term and one-sided first order difference with the direction depending
on the sign of S (so-called {\em upwinding}) is used for the advective term.
For the time-stepping  an implicit Backward-Euler scheme  with a fixed step-size 
$\Delta_t = \Delta_N$ is used.  The following system of
algebraic equations is obtained:
\begin{equation} \label{eq:num_pde_FW} 
G_i^{k+1} =   G_i^{k}  + \Delta_t \left(\frac{x_i(1-x_i)}{2 \rho(t) (\Delta_i)^2} 
\left[G^{k+1}_{i - Edge}-2G_i^{k+1} +G_{i+1}^{k+1}\right]\right) \\ 
\end{equation}
$$
\quad -  S\Delta_t \left(\frac{x_i(1-x_i)}{ \Delta_i } \left[G_i^{k+UwR} - G_i^{k+UwL}\right] \right),
$$
where $Edge = 2$ if $x_i$ is and edge-node and $Edge = 1$ otherwise; 
$UwR=0$,$UwL=-1$ if $S>0$ and $UwR=1$, $UwL=0$ if $S<0$.
The boundary conditions are $G_0^k = g(\Delta_t*k)$ and $G_{N+1}^k = 0$.

Overall the truncation error is of the order $\Delta_N$.  In each time step, in order to solve
the system (\ref{eq:num_pde_FW} ) it is
necessary to invert a sparse matrix with a large condition number ($cN \approx 10^8$), hence
a gain in accuracy that would come from decreasing the time step is offset by
the loss of precision in the inversion of such a matrix.

\section{Model of recent human population growth}
\label{S:hum_pop_growth}

We considered a highly simplified model of the 
history of population sizes of modern humans 
similar to that used by \citet{24} 
and \citet{10} but not requiring 
the assumption of an instantaneous change in 
population size.  We assumed a stable population 
containing $N_0=10,000$ individuals until $150,000$ 
years ago, which we took as time $t=0$.
We assumed a generation time of $25$ 
years. We measured time in units of $2N_0$, so 
the present is at $t=6000/20000=0.3$.  At $t=0$ 
the population began to increase in size 
exponentially at a scaled rate $R=2N_0r$, where 
$r$ is the exponential rate per generation.  We 
assumed additive selection with heterozygous 
fitness $1+s$ relative to {\bf aa} homozygotes, with 
the selection coefficient is also scaled by 
$N_0$: $S=2N_0s$. The value $R=40$ corresponds to a current 
size of $1.63 \times 10^9$.  Reich and Lander 
assumed a current size of $6 \times 10^9$, but that did 
not take account of the fact that the effective 
size of human populations is roughly $1/3$ of the 
census size \citep{25}.  

We assume that the spectrum at $t=0$ is the equilibrium spectrum:
\begin{equation}
\label{eq:XX+3}
g(x,0) = \theta(1-x)
\end{equation}
for $S=0$ and
\begin{equation}
\label{eq:XX+4}
g(x,0) = \theta \frac{e^{2S}\left(1 - e^{-2S(1-x)}\right)}{e^{2S}-1}
\end{equation}
otherwise.  The numerical solutions for $f(x,t)$ at times $t=0$
and $t=0.3$ are plotted in Figures 1--3 for the respective
choices $0,+2,-2$ of the selection parameter $S$ (the mutation parameter
is taken to be $\theta = 1$ -- a different choice of $\theta$ merely rescales
the spectrum).

\begin{center}
{\bf
FIGURE 1 HERE \\
FIGURE 2 HERE \\
FIGURE 3 HERE}
\end{center}

For neutral alleles, the equation for the 0th 
moment of $g(x,t)$, which is half the expected total
heterozygozity, can be solved exactly.  It is of 
interest to separate the solution into two parts, 
one representing alleles present at $t=0$ (old 
alleles) and the other representing alleles that 
arose by mutation after $t=0$ (new alleles).  For 
old alleles,
\begin{equation}
\label{eq:XX+5}
\frac{d\mu_{0,o}}{dt} = - e^{-Rt} \mu_{0,o}
\end{equation}
with initial condition $\mu_{0,o}(0)=\frac{\theta}{2}$ and for new alleles
\begin{equation}
\label{eq:XX+6}
\frac{d\mu_{0,n}}{dt} = \frac{\theta}{2} - e^{-Rt} \mu_{0,n}
\end{equation}
with initial condition $\mu_{0,n}(0)=0$. 
Equations (\ref{eq:XX+5}) and (\ref{eq:XX+6}) can be solved in terms of 
exponential integrals.  With $R=40$, we find 
$\mu_{0,o}(0.3)=0.49  \theta$ and 
$\mu_{0,n}(0.3)=0.15  \theta$, which implies 
that under this model, roughly 30\% of the 
expected heterozygosity at neutral sites is attributable 
to mutations that arose in the past $150,000$ years.

For selected 
alleles ($S \ne 0$), the system for the moments is 
not closed. An approximation to the moments can be
obtained by truncating the system and setting the first
neglected term to it's initial value. Alternatively, the moments
can be computed by first solving for $g(x,t)$ and numerically integrating. Both
approaches present certain numerical difficulties.
The plots in Figure 4--6 were obtained for a sample of size $n=20$ by
 numerically solving the truncated system with $160$ equations using
{\em MATLAB}'s {\tt ode45} (for $S=0$) and {\tt ode15s} (for $S= \pm 2$) routines.
For the neutral case $S=0$ there is a simulation algorithm due to
\citet{13} for approximating the finite spectrum, and we compare that approximation
with the results from the numerical solution of the the system of ODEs in 
Figure 7 for a sample of size $n=40$.  The mutation parameter for Figures 4--7
is taken to be $\theta = 1$.  Again, a different choice of $\theta$ merely rescales
the finite spectrum.  It appears from Figure 1 that the frequency spectrum at time
$t=0.3$ is a decreasing function and hence, from the remark at the end of
Section \ref{S:finite_sample}, the corresponding finite spectrum plotted in
Figure 7 should also be decreasing.  The undulations in the plot are therefore
numerical artifacts.

\begin{center}
{\bf
FIGURE 4 HERE \\
FIGURE 5 HERE \\
FIGURE 6 HERE \\
FIGURE 7 HERE}
\end{center}

\newpage

\begin{figure}[!p]
	\begin{center}
		\includegraphics[width=1.00\textwidth]{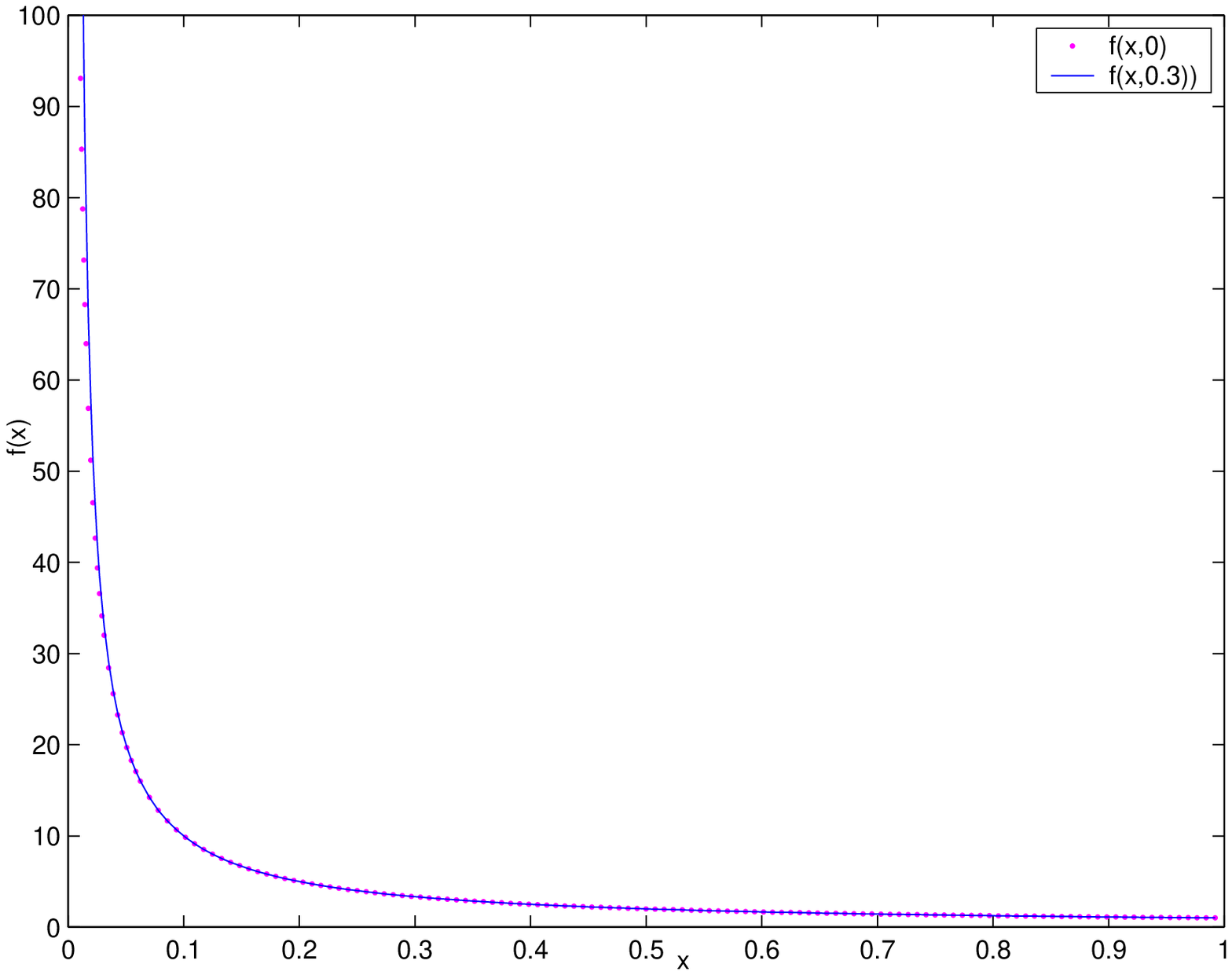}
	\end{center}
	\caption{Frequency spectrum $f(x,t)= g(x,t)/(x(1-x))$ at times
	$t=0$ and $t = 0.3$ with parameter values $R=40$ and $S=0$.
Obtained by numerically integrating the PDE.
The values of $f$ are restricted to the interval $[0,100]$.}
	\label{fig:f_R40_S0}
\end{figure}

\newpage

\begin{figure}[!p]
	\begin{center}
		\includegraphics[width=1.00\textwidth]{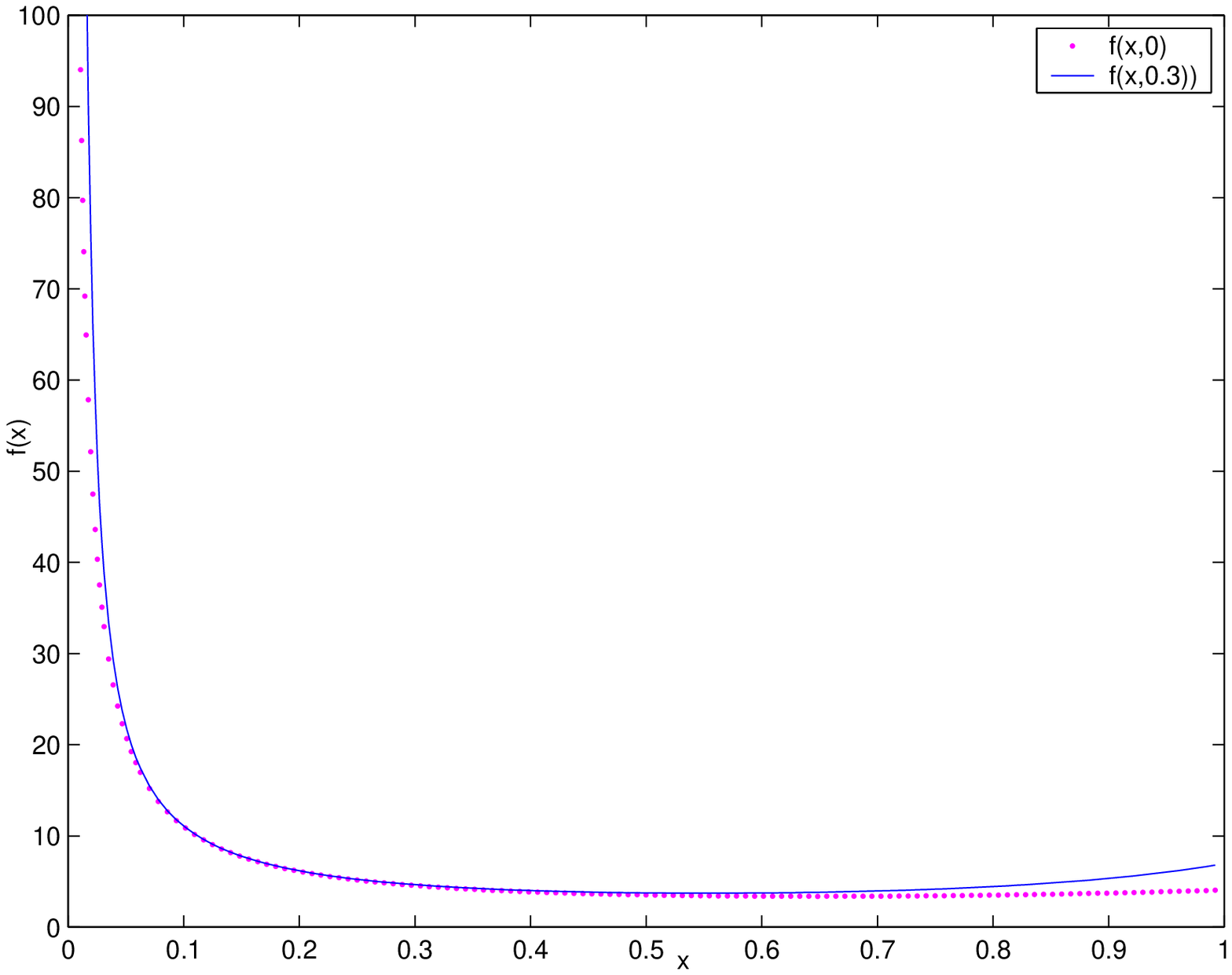}
	\end{center}
	\caption{Frequency spectrum $f(x,t)= g(x,t)/(x(1-x))$ at times
	$t=0$ and $t = 0.3$ with parameter values $R=40$ and $S=+2$.
Obtained by numerically integrating the PDE.
The values of $f$ are restricted to the interval $[0,100]$.}
	\label{fig:f_R40_S2}
\end{figure}

\newpage

\begin{figure}[!p]
	\begin{center}
		\includegraphics[width=1.00\textwidth]{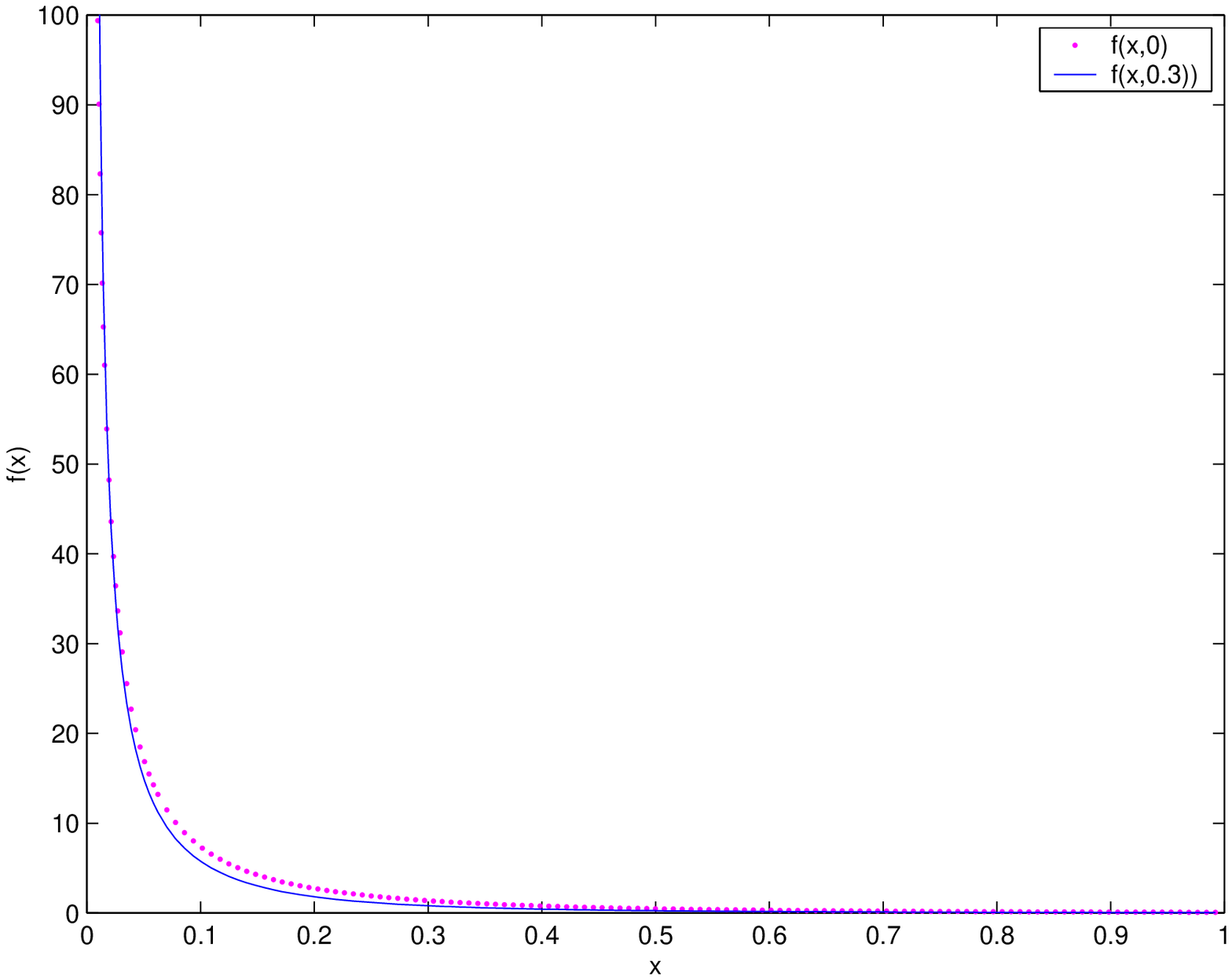}
	\end{center}
	\caption{Frequency $f(x,t)= g(x,t)/(x(1-x))$ at times $t=0$ and $t = 0.3$ with 
	parameter values $R=40$ and $S=-2$;
Obtained by numerically integrating the PDE.
The values of $f$ are restricted to the interval $[0,100]$.}
	\label{fig:f_R40_Smin2}
\end{figure}

\newpage

\begin{figure}[!p]
	\begin{center}
		\includegraphics[width=1.00\textwidth]{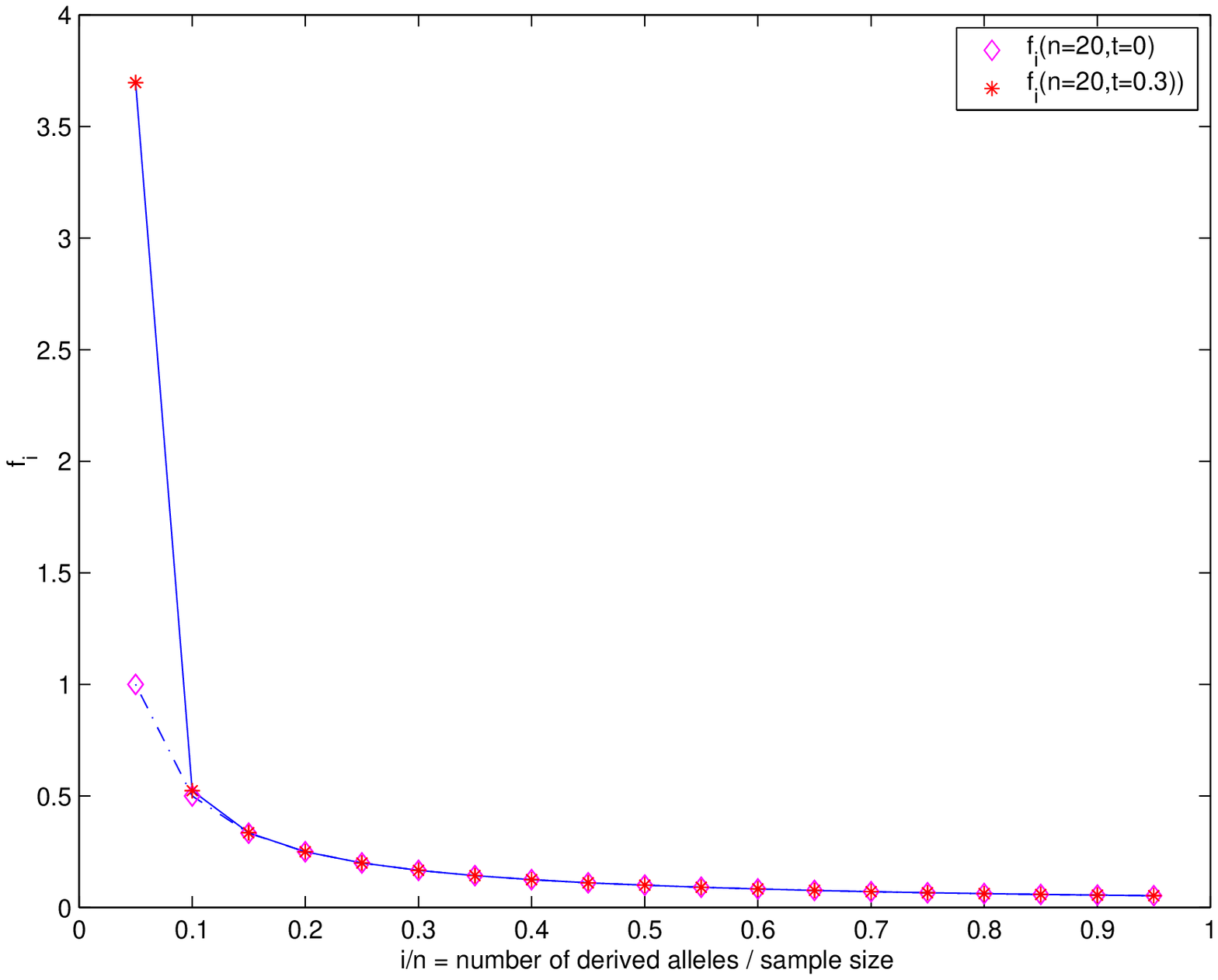}
	\end{center}
	\caption{Finite spectrum for a sample size $n=20$ at times $t=0$ and $t=0.3$
	with parameter values $R=40$ and $S=0$.
Obtained by numerically integrating the system of ODEs for the moments of $g(\cdot,t)$.}
	\label{fig:fi_R40_S0_n20}
\end{figure}

\newpage

\begin{figure}[!p]
	\begin{center}
		\includegraphics[width=1.00\textwidth]{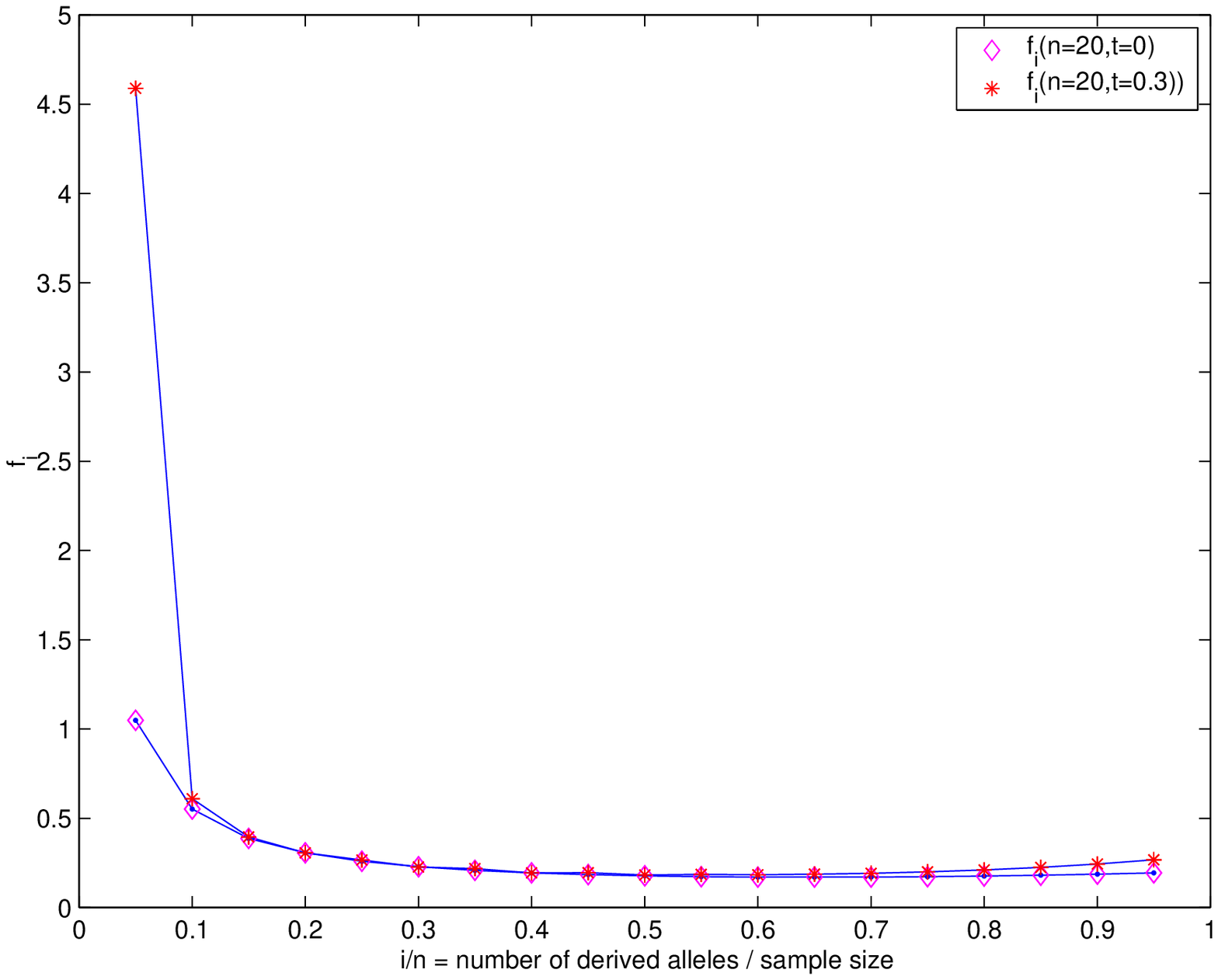}
	\end{center}
	\caption{Finite spectrum for a sample size $n=20$ at times $t=0$ and $t=0.3$
	with parameter values $R=40$ and $S=+2$.
Obtained by numerically integrating the system of ODEs for the moments of $g(\cdot,t)$.}
	\label{fig:fi_R40_S2_n20}
\end{figure}

\newpage

\begin{figure}[!p]
	\begin{center}
		\includegraphics[width=1.00\textwidth]{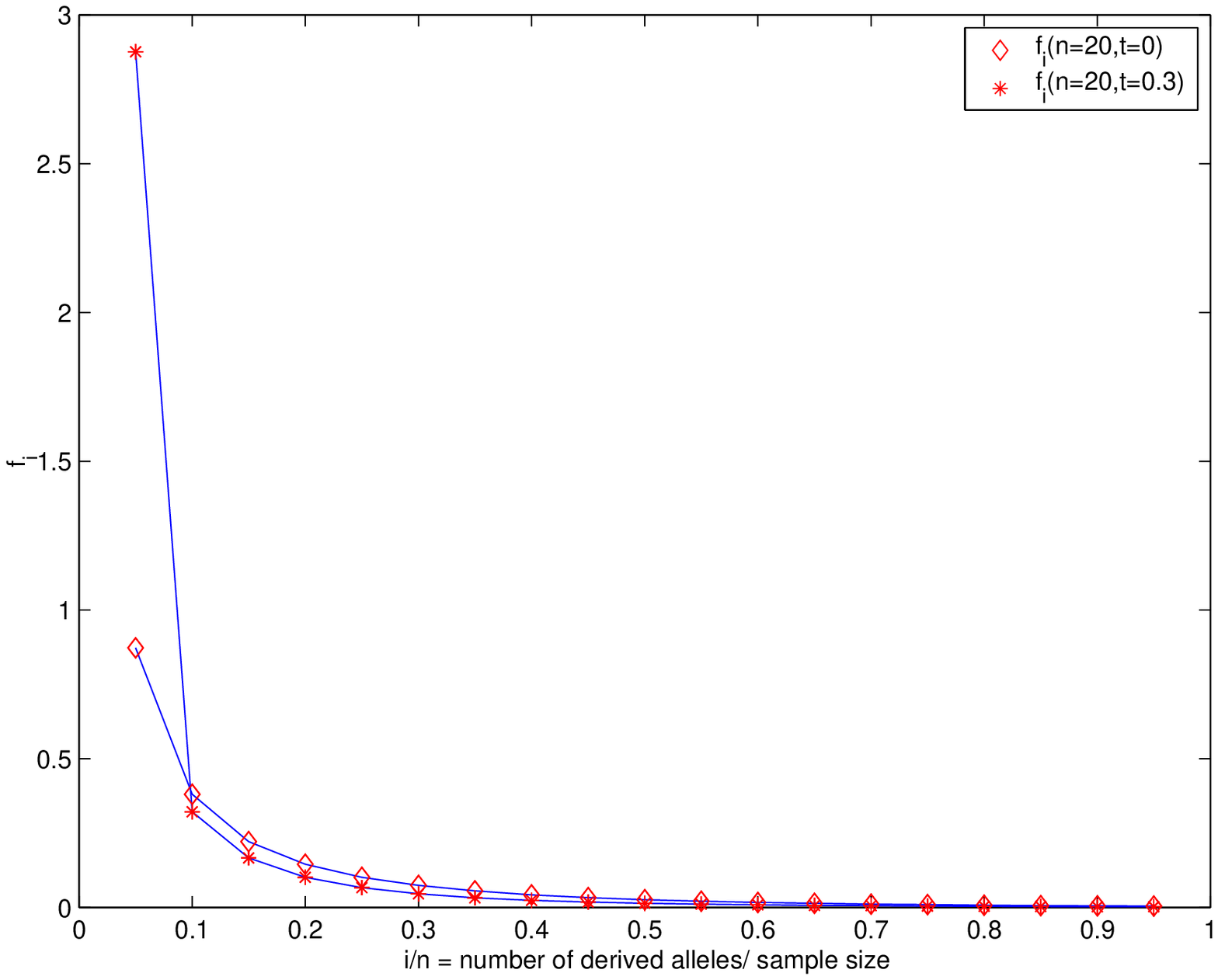}
	\end{center}
	\caption{Finite spectrum for a sample size $n=20$ at times $t=0$ and $t=0.3$
	with parameter values $R=40$ and $S=-2$.
Obtained by numerically integrating the system of ODEs for the moments of $g(\cdot,t)$.}
	\label{fig:fi_R40_Smin2_n20}
\end{figure}

\newpage

\begin{figure}[!p]
	\begin{center}
		\includegraphics[width=1.00\textwidth]{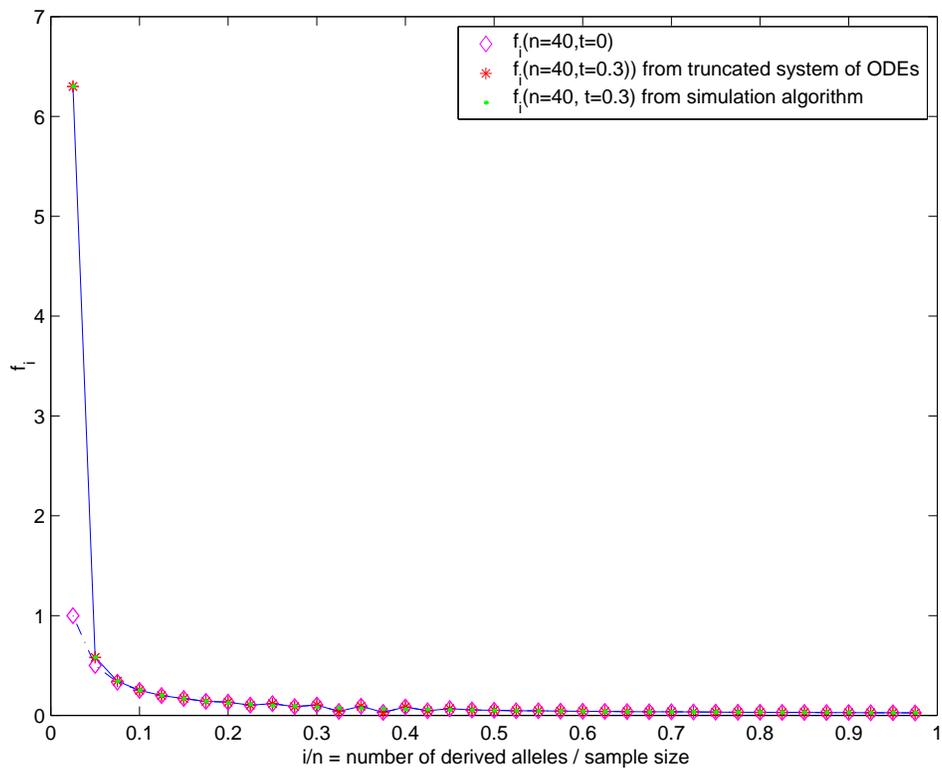}
	\end{center}
	\caption{Finite spectrum for a sample size $n=40$ at times $t=0$ and $t=0.3$
	with parameter values $R=40$ and $S=0$.
Obtained by numerically integrating the system of ODEs for the moments of $g(\cdot,t)$.
Compared at time $t=0.3$ with results from the simulation algorithm of Griffiths and Tavar\'e.}
	\label{fig:fi_R40_S0_n40}
\end{figure}

\end{document}